\documentclass[fleqn,usenatbib]{mnras}

\usepackage{newtxtext,newtxmath}

\usepackage[T1]{fontenc}
\usepackage{ae,aecompl}

\usepackage{graphicx}	
\usepackage{amsmath}	
\usepackage{amssymb}	
\usepackage[usenames, dvipsnames]{color}

\title[A close look at NGC~419 with MUSE+AO]{Cluster kinematics and stellar rotation in NGC~419 with MUSE and adaptive optics\thanks{Based on public data released from the MUSE WFM-AO commissioning observations at the VLT Yepun (UT4) telescope under Programme ID 60.A-9100(G) and 60.A-9100(H).}}

\author[S. Kamann et al.]{
S. Kamann$^{1}$\thanks{E-mail: s.kamann@ljmu.ac.uk}  
N. Bastian$^{1}$,
T.-O. Husser$^{2}$,
S. Martocchia$^{1}$,
C. Usher$^{1}$,\newauthor
M. den Brok$^{3}$,
S. Dreizler$^{2}$,
A. Kelz$^{3}$,
D. Krajnovi\'c$^{3}$,
J. Richard$^{4}$,\newauthor
M. Steinmetz$^{3}$,
P.~M. Weilbacher$^{3}$
\\
$^{1}$Astrophysics Research Institute, Liverpool John Moores University, 146 Brownlow Hill, Liverpool L3~5RF, UK\\
$^{2}$Institut f\"ur Astrophysik, Georg-August Universit\"at, Friedrich-Hund-Platz 1, 37077 G\"ottingen, Germany\\
$^{3}$Leibniz-Institute for Astrophysics (AIP), An der Sternwarte 16, 14482 Potsdam, Germany \\
$^{4}$Univ Lyon, Univ Lyon1, Ens de Lyon, CNRS, Centre de Recherche Astrophysique de Lyon UMR5574, F-69230, Saint-Genis-Laval, France \\
}

\date{Accepted XXX. Received YYY; in original form ZZZ}

\pubyear{2018}

\begin{document}
\label{firstpage}
\pagerange{\pageref{firstpage}--\pageref{lastpage}}
\maketitle

\begin{abstract}
We present adaptive optics (AO) assisted integral-field spectroscopy of the intermediate-age star cluster NGC~419 in the Small Magellanic Cloud. By investigating the cluster dynamics and the rotation properties of main sequence turn-off stars (MSTO), we demonstrate the power of AO-fed MUSE observations for this class of objects. Based on $1\,049$ radial velocity measurements, we determine a dynamical cluster mass of $1.4\pm0.2\times10^5\,{\rm M_\odot}$ and a dynamical mass-to-light ratio of $0.67\pm0.08$, marginally higher than simple stellar population predictions for a Kroupa initial mass function. A stacking analysis of spectra at both sides of the extended MSTO reveals significant rotational broadening. Our analysis further provides tentative evidence that red MSTO stars rotate faster than their blue counterparts. We find average $V\sin i$ values of $87\pm16\,{\rm km\,s^{-1}}$ and $130\pm22\,{\rm km\,s^{-1}}$ for blue and red MSTO stars, respectively. Potential systematic effects due to the low spectral resolution of MUSE can reach $30\,{\rm km\,s^{-1}}$ but the difference in $V\sin i$ between the populations is unlikely to be affected.
\end{abstract}



\section{Introduction}


One peculiar feature of young and intermediate age (<2 Gyr) massive clusters in the LMC/SMC is that they display extended MSTOs.  The origin of this feature has been the subject of much debate in recent years, with the main explanations being either age spreads of 10s to 100s of Myr within the clusters \citep[e.g.][]{Mackey2008,Goudfrooij2014} or a spread in stellar rotation rates \citep{Bastian2009,Niederhofer2015,Brandt2015b}, or both \citep{Dupree2017}.  If massive clusters were shown to host such large age spreads, our view of the formation and early evolution of such objects would need to be radically redefined \citep{PortegiesZwart2010,Bastian2017}.  If, on the other hand, massive clusters hosted stars with a wide range of rotational velocities (from non-rotating to near-critical rotation), models predict a strong correlation between the colour and the projected rotation velocity of a star \citep[e.g.][]{Georgy2014}. In \citet{Bastian2018}, we recently showed that the predicted correlation is in good agreement with the measured rotation velocities in the 800~Myr old open cluster NGC~2818.  

Additional questions remain about the global kinematics of massive clusters.  Recent studies have suggested that massive clusters display significant global rotation, at young \citep{Henault-Brunet2012}, intermediate \citep{Davies2011q,Mackey2013b} and old ages \citep[e.g.][]{Kamann2018}.  This raises questions about the initial range and the age evolution of V/$\sigma$.  In order to address this, large samples of clusters at all evolutionary stages need to be surveyed.

Finally, massive stellar clusters routinely serve as calibration sources in population synthesis studies because their integrated properties (e.g., mass-to-light ratios and their integrated spectra/colours) can be linked directly to the known underlying (resolved) stellar population.  NGC~419, at an age of $\sim1.5$~Gyrs \citep{Glatt2008}, lies at an age where integrated properties are particularly insecure, due to the uncertainties in modelling the stellar evolutionary phases of stars that are dominating the light \citep[e.g.][]{Girardi2013a}.  By measuring the mass-to-light ratio in such a cluster, strict constraints on simple stellar population models can be obtained.


In this work we demonstrate how the new MUSE capabilities \citep[e.g.][]{Leibundgut2017} can help answering the aforementioned questions, even with relatively short exposure times of around one hour. To this aim, we exploit MUSE observations of NGC~419, performed during the commissioning of its new adaptive optics (AO) system.  NGC~419 has one of the most extended MSTOs observed to date and we use the data to measure the stellar rotation rates along the MSTO. In addition, we infer the global kinematics of the cluster and measure the dynamical mass-to-light ratio.

%

\section{Observations \& data reduction}

The data were taken on the night of 2017-07-14 during the second commissioning run of AOF/GALACSI \citep{LaPenna2016,Arsenault2017c}, the new AO system for MUSE \citep{Bacon2010}. In total, 8 exposures of 600~s each were taken. To verify the image quality improvement caused by the AO, the first four exposures were alternated between open and closed AO loop. The remaining four exposures were taken with closed loop.

\begin{figure}
 \includegraphics[width=\columnwidth]{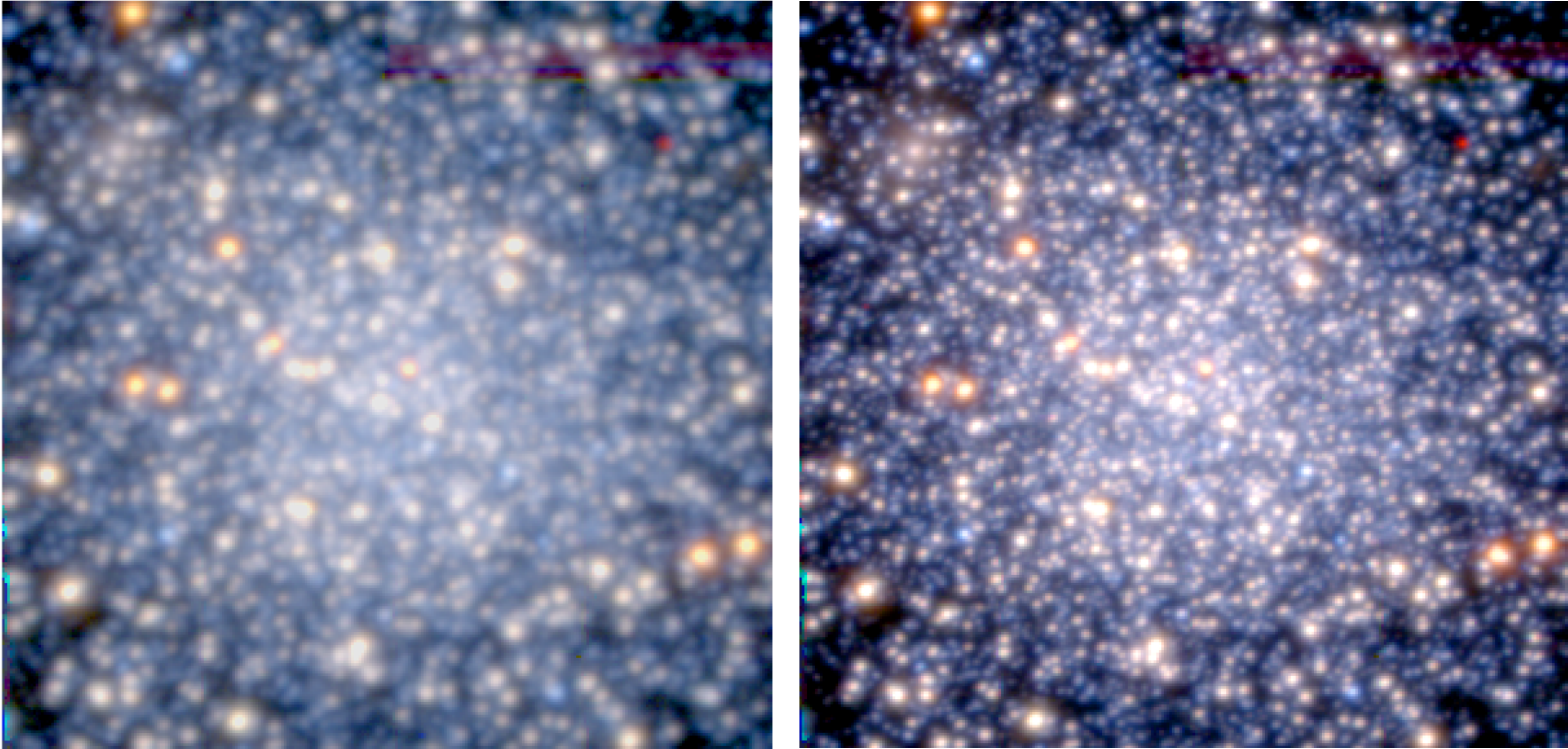}
 \caption{SDSS {\it gri} colour images of NGC~419, created from MUSE cubes without ({\it left}) and with ({\it right}) AO support, observed back-to-back. Each image is $1\arcmin\times1\arcmin$ in size, the halflight radius of NGC~419 is $28\arcsec$ \citep{Glatt2009}. North is up, east is left.}
 \label{fig:fov_noao_ao}
\end{figure}

We employed version 2.1.1 of the official MUSE pipeline \citep{Weilbacher2014} to reduce the data. It differs from previous versions mainly by the added support for the AO mode and we refer the interested reader to{\it ,} e.g., \citet{Kamann2018}, for details on the data reduction process. From each science exposure, we created a flux calibrated data cube with the night sky subtracted. In Fig.~\ref{fig:fov_noao_ao}, we compare re-constructed images from MUSE data cubes observed with and without AO support. The improvement in image quality delivered by GALACSI is nicely visible. The extremely red source visible in the north western corner of both cubes is an asymptotic giant star with extreme mass loss, originally discovered by \citet{Tanabe1997}.



We combined the six AO exposures into our final data cube. To account for clouds passing during the observations, we adapted the exposure time of each cube to its flux level relative to the first exposure.


\section{Data analysis}
\label{sec:analysis}
We used our dedicated extraction code for integral field data described in \citet{Kamann2013} to extract stellar spectra from the data cube of NGC~419. As reference photometry catalogue, we used the data from the study of \citet{Martocchia2017}. It includes HST/ACS and WFC3 magnitudes ranging from the UV to the near infrared. Some bright stars that were found to be missing in the HST data because of saturation were added using coordinates and magnitudes from Gaia data release one \citep{GaiaCollaboration2016,GaiaCollaboration2016a}.

We ran the source extraction on a combined cube created from the six AO exposures and on all individual exposures. The latter allowed us to get an idea of how the AO affects the PSF. For the two exposures shown in Fig.~\ref{fig:fov_noao_ao}, the FWHM of the Moffat profile improved from $0.74\arcsec$ ($0.89\arcsec$) to $0.35\arcsec$ ($0.54\arcsec$) at $8\,000\,$\AA\, ($5\,000\,$\AA) when using the AO. At the same time we observed a decrease in $\beta$ (which measures the kurtosis of a Moffat profile) from $2.8$ ($3.6$) to $1.8$ ($2.2$). This agrees with the general trend that the enhanced AO resolution comes at the price of pronounced PSF wings. Still, it is remarkable that even at the blue end of the wavelength range, the AO significantly improves the spatial resolution of MUSE. This also affects the number of resolvable stars. The number of extracted spectra increased by from $2\,254$ in the non-AO exposure to $5\,015$ in the AO exposure. In addition, the average S/N of spectra extracted from both exposures improved by a factor of $1.7$.

\begin{figure*}
 \includegraphics[width=\textwidth]{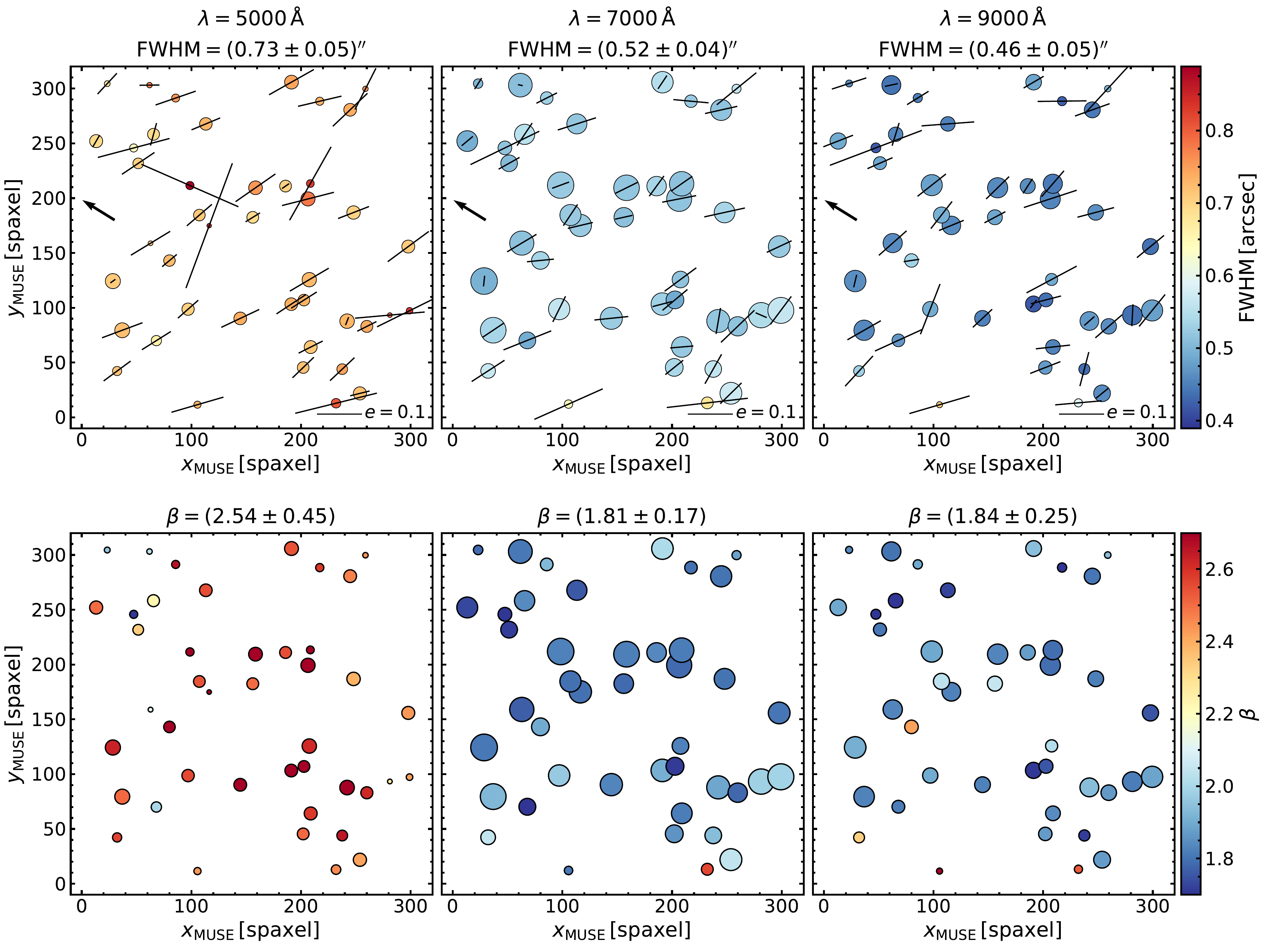}
 \caption{Variation of the MUSE AO-PSF with wavelength and across the field of view. The panels show the values of the FWHM ({\it top}) and $\beta$ ({\it bottom}) obtained in single PSF fits of bright stars at $5\,000\,$\AA\ ({\it left}), $7\,000\,$\AA\ ({\it middle}), $9\,000\,$\AA\ ({\it right}) as a function of the stars positions in the MUSE cube. Mean values and standard deviations of the parameters are provided at the top of each panels. The sizes of the circles scale with the S/N of the stars at the provided wavelengths. In the top panel, black lines indicate the ellipticities and the position angles of the best-fit PSF models. The length of the lines scales linearly with ellipticity as indicated in the bottom right corner of each panel. The direction towards the natural guide star is indicated by a black arrow.}
 \label{fig:psf_variability}
\end{figure*}

We further used the combined data cube to inspect the spatial behaviour of the PSF. AO observations often suffer from a degradation of the image quality with increasing distance to the natural guide star. As GALACSI additionally uses four laser guide stars that are placed around the MUSE field of view, spatial variations are expected to be small. Still, stronger variations could potentially affect the source extraction as it assumes the PSF to be constant across the field of view. We selected $\sim50$ reasonably bright and isolated stars distributed across the field of view. Each star was fitted with a single elliptical Moffat profile. In order to minimize the impact of nearby stars, the fit was performed iteratively and after each iteration nearby stars were subtracted using the current PSF model. The results of this analysis are summarized in Fig.~\ref{fig:psf_variability}. Overall, the variation of the PSF across the field of view seems to be small. The star-to-star scatter in the FWHM is of the order of 10 per cent, while it seems to be larger for $\beta$. However, a significant fraction of the scatter can be attributed to stars with low S/N and hence poorly constrained fits. No systematic variation is found with respect to the distance or position angle towards the natural guide star.

We find the PSF to be slightly elliptical, with a mean fitted ellipticity of $e=0.07\pm0.04$. While the position angles of the semimajor axes seem to be well aligned, they are not directed towards the guide star as one would expect in AO observations that only use a natural guide star (cf. Fig.~\ref{fig:psf_variability}). We conclude that the AO does not introduce significant PSF variations across the field of view of MUSE that could affect the extraction of the stellar spectra.

\begin{figure}
 \includegraphics[width=\columnwidth]{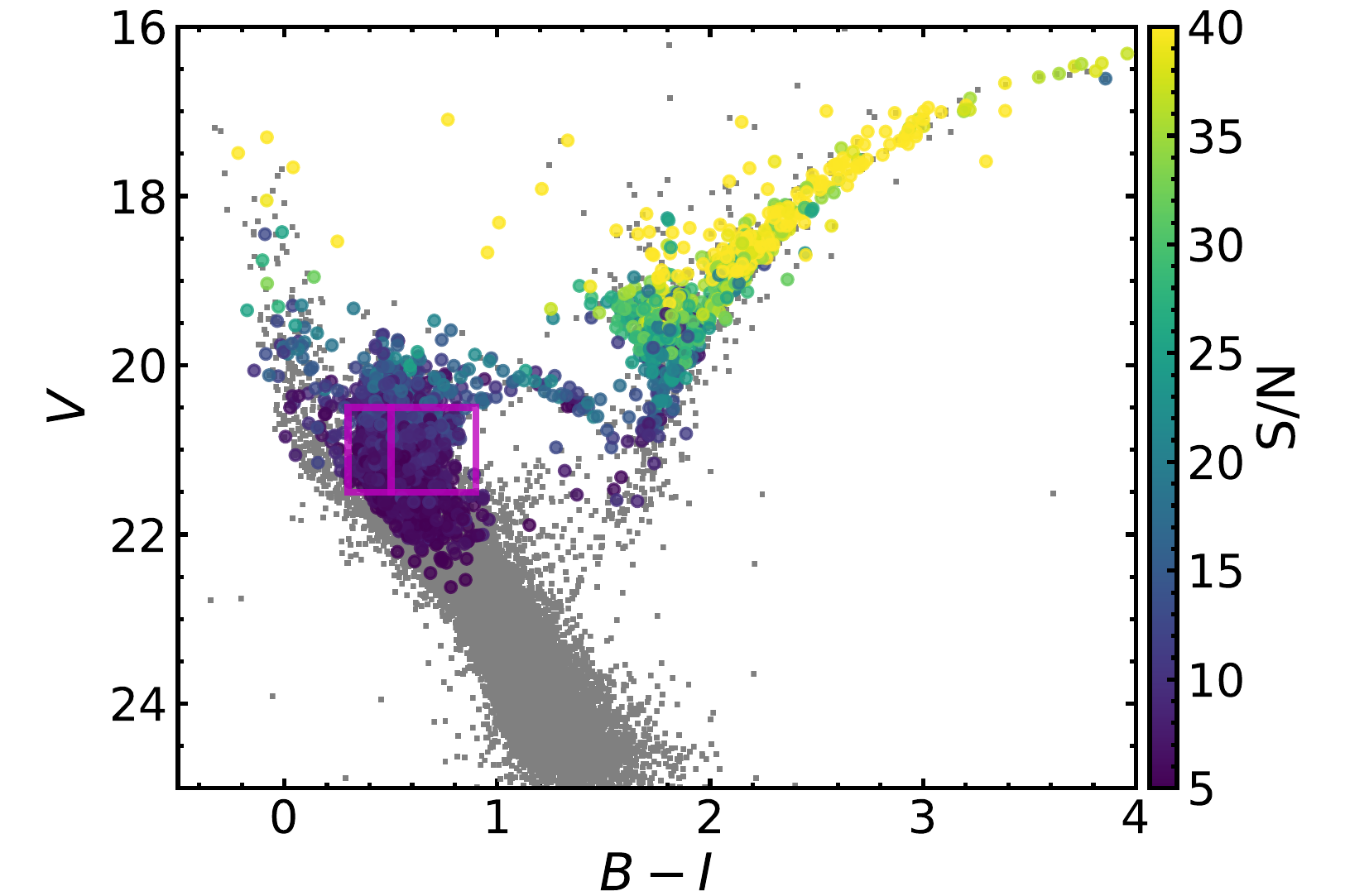}
 \caption{HST ($B-I$, $V$) colour magnitude diagram of NGC~419. Stars for which spectra with ${\rm S/N}>5$ were extracted from the MUSE data are colour coded according to the spectral S/N as indicated by the colour bar to the right. The purple boxes indicate the selection boxes for the red and blue MSTO stars, respectively.}
 \label{fig:cmd}
\end{figure}

In total, we extracted spectra for $5\,538$ stars over a wide range of signal-to-noise ratios (S/Ns). In the following, we will only use the spectra that were extracted with a ${\rm S/N}>5$ per pixel (averaged over the full spectrum), which we consider the minimum requirement for any analysis. Applying this cut results in a sample of $3\,321$ stars with useful spectra (60\% of the parent sample). Their distribution across the colour-magnitude diagram of NGC~419 is shown in Fig.~\ref{fig:cmd}. The faintest stars with spectra passing the S/N cut are situated about $2\,{\rm mag}$ below the MSTO while spectra of stars at the MSTO were extracted with a ${\rm S/N}\sim10$.

\section{Dynamical cluster mass}
\label{sec:dynamics}

To obtain radial velocities for our spectra, we cross correlated them against synthetic PHOENIX templates from the library of \citet{Husser2013}. As a consequence of the broad range in S/N of the extracted spectra, however, not every spectrum allows for a reliable radial velocity measurement. Besides the S/N, we also used the height of the normalised cross-correlation peak  \citep[parametrised by the value $r$ as defined by][]{Tonry1979} and only considered results from spectra where ${\rm S/N} > 5$ and $r>4$ were simultaneously fulfilled.

To investigate the internal dynamics of star clusters, it is further crucial to properly calibrate the uncertainties of the measured radial velocities. A systematic over- or underestimation of the uncertainties would result in a bias in the measured velocity dispersion. As outlined in Sect.~\ref{sec:analysis}, besides the combined cube we also analysed the individual AO exposures. They provided us with multiple independent spectra for the majority of the stars in our sample. Following our previous work \citep[e.g.][]{Husser2016}, we scaled the uncertainties returned by the cross-correlation routine for the individual spectra such that they were consistent with the differences in the radial velocities measured using spectra of the same stars. To account for possible trends with the S/N, the calibration was carried out in five S/N bins. The correction factors we found were around 10 per cent. For each star, we finally obtained its radial velocity and the associated uncertainty as the inverse-variance weighted mean and variance of the individual measurements. As expected, the measurement uncertainties strongly scale with the brightnesses and spectral types of the stars. Down the red giant branch, the average velocity uncertainties increase from $1\,{\rm km\,s^{-1}}$ to $7\,{\rm km\,s^{-1}}$. Along the main sequence, the stars are hotter and hence their spectra show fewer and broader lines. Accordingly, their radial velocities are less well defined, with typical uncertainties ranging from $15\,{\rm km\,s^{-1}}$ at the tip of the MSTO to $35\,{\rm km\,s^{-1}}$ for the faintest stars in our sample. In view of the expected velocity dispersion of NGC~419 of $5\,{\rm km\,s^{-1}}$, we excluded all stars with uncertainties $>10\,{\rm km\,s^{-1}}$ from the analysis of the cluster dynamics. This left us with $1\,049$ measurements, obtained mainly from stars on the red giant branch.

The dynamics of the cluster was analysed using the same approach as in \citet{Kamann2018}. It finds the maximum-likelihood solution for the intrinsic dispersion and the rotation field of the cluster given the data. Further, we used the MCMC approach of \citet{Foreman-Mackey2013} to obtain confidence intervals for the measured quantities. The formal uncertainties provided in the following are the 16th and 84th percentiles of the parameter distributions obtained from the MCMC walkers.

From an analysis of the full sample, we obtained a systemic radial velocity of NGC~419 of $v_{\rm sys}=190.5\pm 0.2\,{\rm km\,s^{-1}}$ which is in good agreement with $188.3\pm0.9\,{\rm km\,s^{-1}}$, obtained by \citet{Dubath1997} from an integrated spectrum. The analysis further yielded a velocity dispersion of $3.1\pm 0.2\,{\rm km\,s^{-1}}$ and a rotation amplitude of $0.7\pm 0.2\,{\rm km\,s^{-1}}$. The rotation axis is aligned approximately in north-south direction, with a position angle of $13\pm17\,{\rm degrees}$. Only one per cent of the MCMC samples have a rotation amplitude below the uncertainty quoted above, indicating a low probablitity for a non-rotating cluster.

\begin{figure}
 \includegraphics[width=\columnwidth]{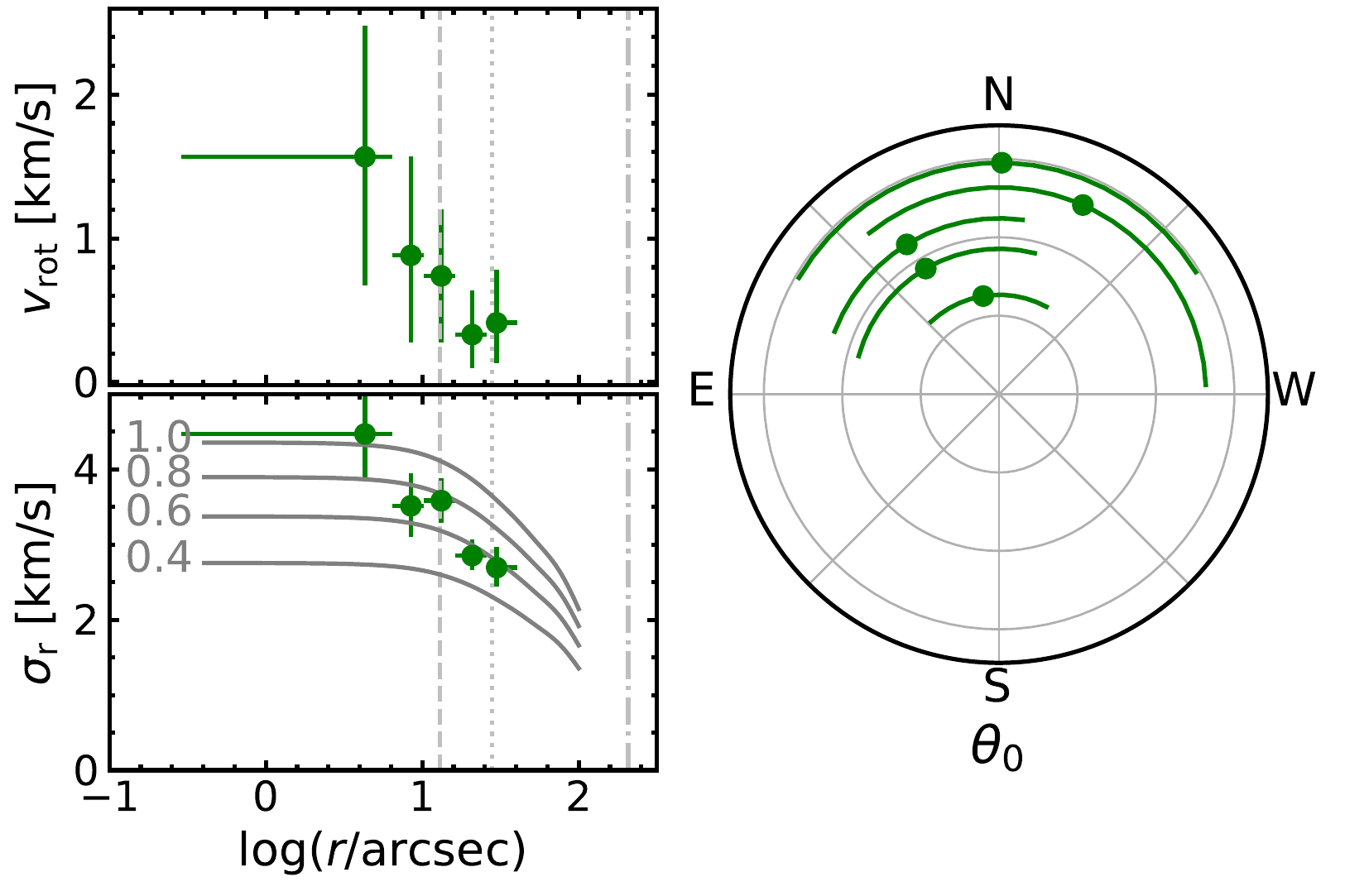}
 \caption{Internal dynamics of NGC~419 as determined from the MUSE data. The left panels show the amplitude of the rotation field ({\it top}) and velocity dispersion ({\it bottom}) as a function of distance to the cluster centre, the behaviour of the axis angle of the rotation field is shown in the right panel. The dispersion profile is compared to the predictions of spherical isotropic Jeans models with different $V$-band mass-to-light ratios, ranging from $0.4\times$ solar to $1.0\times$ solar (in steps of $0.2$). Further, we show the core (dashed line), half-light (dotted line), and truncation radius (dash-dotted line) from the structural analysis of \citet{Glatt2009}.}
 \label{fig:dynamics}
\end{figure}

For further analyses, we split our data into radial bins (each containing at least 100 measurements and covering a width of $\delta \log (r/1\arcsec) \geq 0.2$) and analysed them independently. Figure~\ref{fig:dynamics} shows the results obtained for the individual bins. It is reassuring that for each of the radial bins, we obtain an angle of the rotation axis that is consistent with the value of $13\pm17\,{\rm degrees}$ measured for the full sample. The rotation velocity seems to increase towards the centre of NGC~419, albeit the significance of this observation is low. If confirmed, this behaviour would be surprising. In their analysis of the LMC cluster NGC~1846, which has a similar age to NGC~419, \citet{Mackey2013b} used the violent relaxation model of \citet{Lynden-Bell1967} to describe the rotation profile. It predicts an increase of the rotation velocity with distance to the cluster centre inside the half-light radius. Most rotating globular clusters also show this behaviour, \citep[e.g.,][]{Bianchini2013,Fabricius2014,Kamann2018}. In addition, dedicated models for rotating globular clusters \citep[e.g.][]{Varri2012} predict an increase in the rotation velocity with radius as well, up to a maximum that is reached at a few half-light radii and beyond which the rotation velocity steadily decreases. However, our data lacks the spatial coverage for a detailed comparison to such models.

The velocity dispersion profile depicted in Fig.~\ref{fig:dynamics} further allows us to determine the mass-to-light ratio (M/L) and the dynamical mass of NGC~419. To this aim, we performed a multi-Gaussian expansion \citep[MGE, see][]{Cappellari2002} of the surface brightness profile of NGC~419. We used the parametric representation of the surface brightness profile obtained by \citet{Carvalho2008k} as input. Then we calculated isotropic, spherical Jeans models with different (constant) M/Ls using the code of \citet{Cappellari2008}. The velocity dispersion profiles predicted by Jeans models with representative M/Ls are included in Fig.~\ref{fig:dynamics}. To find the most likely M/L given our data, we used the same maximum likelihood approach adopted by \citet{Merritt1993} and \citet{Gerssen2002}. It calculates a likelihood for each model under the assumption that each measured radial velocity is drawn from a Gaussian distribution which has a variance equal to the sum of squares of the measurement uncertainty and the velocity dispersion predicted by the model for the position of the star. Hence, it requires no binning of the data; the bins shown in Fig.~\ref{fig:dynamics} are used for displaying purposes only. This analysis yields a M/L in the $V$-band in solar units of $0.67\pm0.06$, where the $1\sigma$ confidence interval has been obtained using a likelihood ratio test. It does not account for uncertainties in the surface brightness profile determination. To get an idea of the uncertainty in the surface profile determination, we also used the \citet{King1962} profile determined by \citet{Goudfrooij2014} as input to our Jeans modelling, which reduced the M/L of the best matching model by $0.05$. The difference can be ascribed to an excess of light in the \citet{Goudfrooij2014} profile. Its numerical integration yields $m_{\rm V}=10.24$ compared to $m_{\rm V}=10.35$ for the \citet{Carvalho2008k} profile. We add the offset in M/L of $0.05$ in quadrature to our uncertainties which gives a final result result of $0.67\pm0.08$. Our value is only slightly above the predictions for simple stellar populations based on a \citet{Kroupa2002} IMF at an age of $1.5\,{\rm Gyr}$ \citep{Glatt2008}. We obtain predictions of $0.48\pm0.05$ using MILES \citep{Vazdekis2010}, $0.50\pm0.06$ using FSPS \citep{Conroy2009,Conroy2010}, $0.62\pm0.04$ using the 2016 version of the \citet{Bruzual2003} models, and $0.53\pm0.03$ using the models of \citet{Maraston2005}.


As mentioned above, numerical integration of the profile by \citet{Carvalho2008k} results in an apparent cluster brightness of $m_{\rm V}=10.35$. In combination with a distance modulus of NGC~419 of $(m-M)=18.85\pm0.03$ \citep{Goudfrooij2014}, we obtain a dynamical cluster mass of $1.4\pm0.2\times10^5\,M_{\rm \odot}$. This is lower than the value of $2.4\pm0.4\times10^5\,M_{\rm \odot}$ determined by \citet{Goudfrooij2014}, who converted the measured light to mass by adopting an simple stellar population (SSP) model. However, \citet{Goudfrooij2014} assumed a \citet{Salpeter1955} IMF in their stellar population synthesis and mentioned that the usage of a \citet{Kroupa2002} IMF instead would lower their masses by a factor of $\sim1.6$, explaining most of the discrepancy to our measurement. Hence our study suggests that a \citet{Kroupa2002} IMF yields a better representation of the intrinsic IMF of NGC~419.

We note that the dispersion measurements in Fig.~\ref{fig:dynamics} suggest a slightly steeper radial profile than the Jeans models. The reason for this is currently unclear, as anisotropy, a radially varying M/L, or binary stars could all affect our analysis. However, a detailed exploration of all mentioned effects is not possible with the current data.

\section{Stellar rotation and the extended MSTO}
\label{sec:rotation}

Figure~\ref{fig:cmd} shows that our sample contains a good fraction of stars around the main sequence turn-off of NGC~419. According to the oldest models in the grid of \citet[][$1\,{\rm Gyr}$]{Georgy2014}, one would expect the projected rotation velocity $V\sin i$ to gradually increase with colour if the extended MSTO was caused by stellar rotation. Hence, we defined two selection boxes, covering the red and the blue side of the MSTO as indicated in Fig.~\ref{fig:cmd}. The division between the two boxes at $B-I=0.5$ was chosen such that each of them contained a comparable number of stars from our sample ($\sim250$).

\begin{figure}
 \includegraphics[width=\columnwidth]{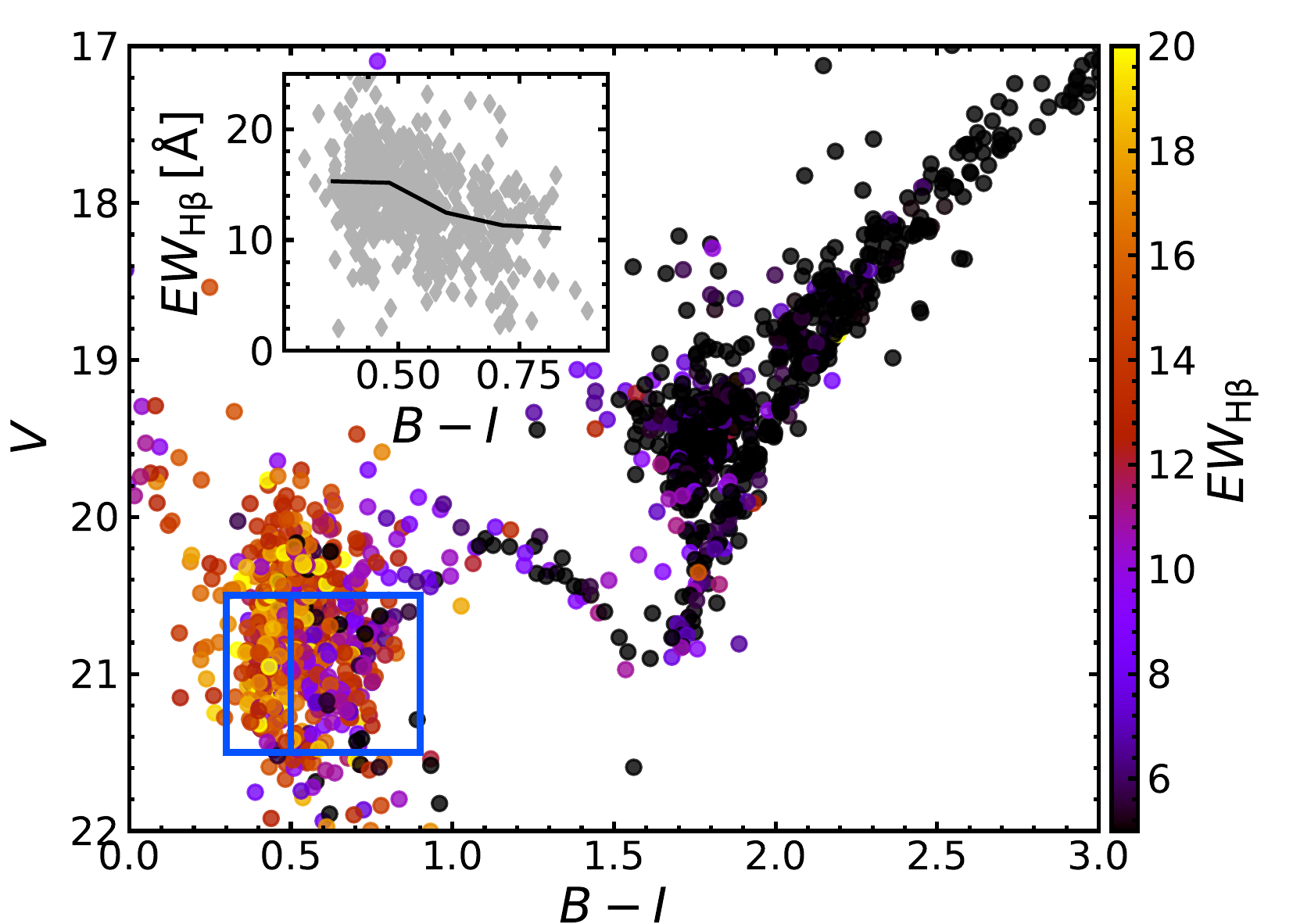}
 \caption{CMD of NGC~419, colour-coded by the measured equivalent width of \ion{H}{$\beta$}. The inset shows the equivalent width variation as a function of $(B-I)$-colour for stars at the main sequence turn-off of NGC~419 (highlighted by the blue boxes), with the running median indicated by a black solid line.}
 \label{fig:hbeta_ew}
\end{figure}

We first determined the equivalent width of the \ion{H}{$\beta$} line in the spectra of the selected stars. Given the comparable age and evolutionary status of the stars, \ion{H}{$\beta$} should be sensitive to changes of the effective temperature. As visible in Fig.~\ref{fig:hbeta_ew}, we find the equivalent width of \ion{H}{$\beta$} to decrease with increasing $B-I$ colour. The mean values for the two sub-samples are $15.2\pm0.3$\AA\ and $12.5\pm0.3$\AA, respectively. At about $7\,500\,{\rm K}$ (the expected temperature for MSTO stars in NGC~419), the difference in equivalent width corresponds to a difference in effective temperature of $300$--$400\,{\rm K}$.


Measuring stellar rotation rates is challenging because of the low spectral resolution of MUSE and the relatively low S/N of the MSTO stars in these commissioning data. To circumvent the latter problem, we made use of our large sample size and averaged the normalised spectra of the stars that fell in either of the two selection boxes highlighted in Fig.~\ref{fig:cmd}. Uncertainties for the combined spectra were determined via bootstrapping of the input spectra. The different radial velocities of the stars we consider will spuriously broaden the stacked spectra. While our analysis of Sect.~\ref{sec:dynamics} provided us with radial velocities also for the MSTO stars, their uncertainties are considerable. For this reason, we did not correct the individual spectra for the measured radial velocities, but only for the systemic velocity of NGC~419, $v_{\rm sys}=190.5\,{\rm km\,s^{-1}}$, determined in Sect.~\ref{sec:dynamics}. The intrinsic velocity dispersion of NGC~419 of $\lesssim4\,{\rm km\,s^{-1}}$ is much lower than the expected rotational velocities, hence the radial velocities of the cluster stars should not affect the following analyses. We further verified that the contamination from SMC field stars in our selection boxes is low, $\lesssim10\%$, by using the same method described in \citet{Martocchia2017}. The influence of field stars was further reduced by only accepting stars with a reliable velocity measurement within $30\,{\rm km\,s^{-1}}$ of the systemic velocity of NGC~419.

\begin{figure*}
 \includegraphics[width=\textwidth]{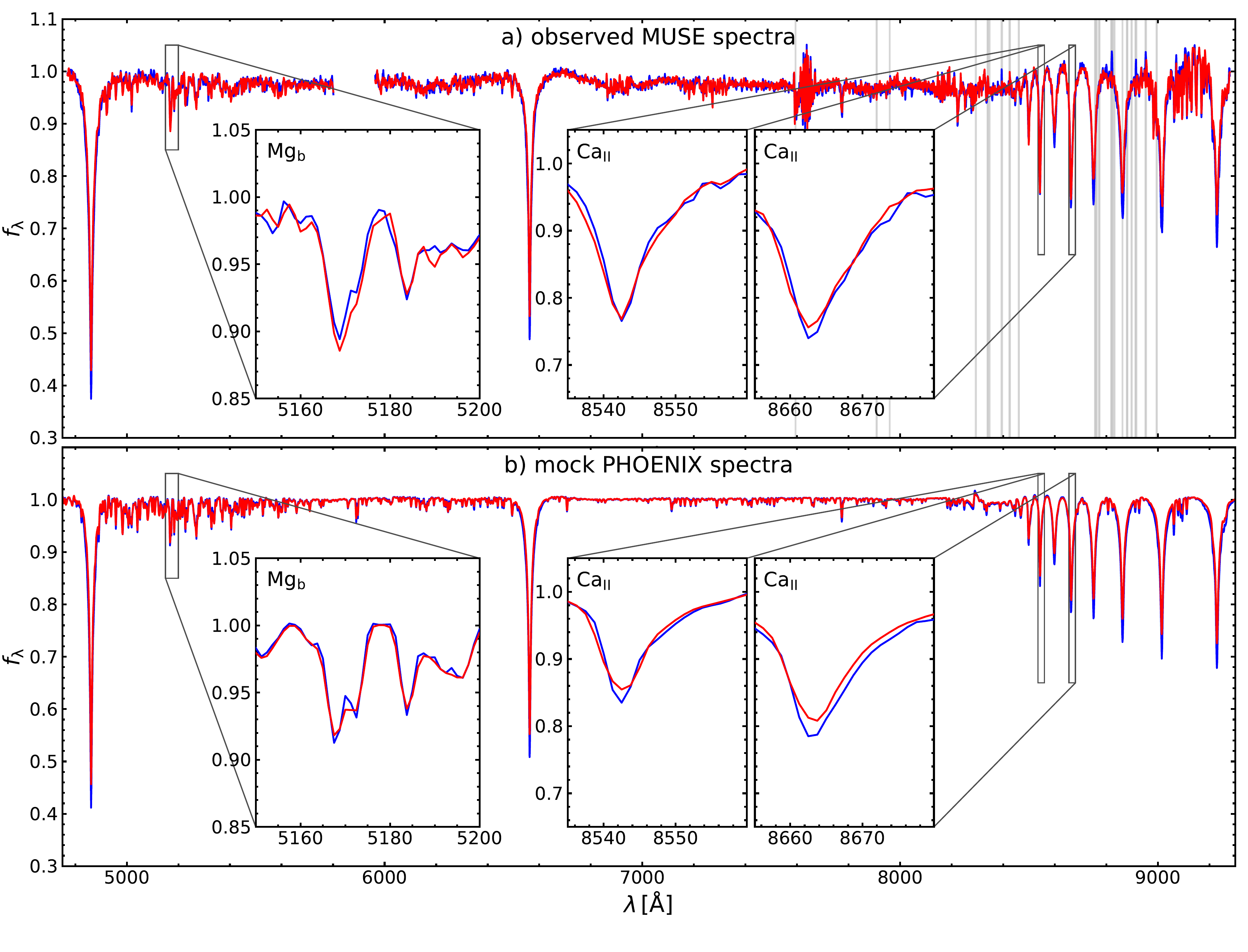}
 \caption{\textit{(Top)} Comparison of combined MUSE spectra of MSTO stars in NGC~419 with $B-I < 0.5$ (blue) and $>0.5$ (red), created by combining the (normalised) spectra extracted for stars in the boxes highlighted in Fig.~\ref{fig:cmd}. The insets in the upper panel shows a zoom to the \ion{Mg}{b} triplet and to the strongest lines of the \ion{Ca}{ii} triplet. Gray-shaded areas indicate wavelength ranges affected by strong telluric emission lines. \textit{(Bottom)} The same for synthetic PHOENIX templates with stellar parameters comparable to those obtained for the observed MSTO spectra. The spectra were transformed to the MUSE spectral resolution and sampling and broadened assuming $V\sin i$ values of $87\,{\rm km\,s^{-1}}$ (blue) and $130\,{\rm km\,s^{-1}}$ (red), respectively.}  
 \label{fig:msto_spectra}
\end{figure*}

The two combined spectra are depicted in the upper panel of Fig.~\ref{fig:msto_spectra}. The S/N determined in both cases is $\sim180$. Both spectra are dominated by the strong \ion{H}{$\beta$} and \ion{H}{$\alpha$} lines as well as the Paschen series of hydrogen. The insets in the upper panel of Fig.~\ref{fig:msto_spectra} zoom in on the ${\rm Mg_b}$ triplet and the strongest lines of the \ion{Ca}{ii} triplet. These lines are the strongest apart from the neutral hydrogen lines, over which they have the advantage of a much lower line width. A visual comparison of the two spectra does not reveal a clear indication that the red MSTO spectrum has broader lines (corresponding to a higher rotation velocity) than the blue MSTO spectrum. However, one has to keep in mind the rather low spectral resolution of MUSE and the temperature difference between the red and blue MSTO stars, as the higher temperature of the latter stars may counterbalance a higher rotation velocity of the former. Therefore, any conclusion on the rotation velocities across the MSTO requires a more detailed analysis, which we carry out below.

In the absence of isolated lines suited for direct rotation measurements, we obtained rotation velocities for the combined spectra as follows. First, the data were analysed with the spectral synthesis code of \citet{Husser2016} under the assumption that $V\sin i=0$, yielding the stellar parameters listed in Table~\ref{tab:stellar_parameters}.\footnote{As explained in \citet{Husser2016}, we denote metallicities measured from MUSE spectra by $[{\rm M/H}]$ instead of $[{\rm Fe/H}]$ to account for the fact that the values are based on a fit of the full spectrum instead of individual iron lines.} The temperature difference of $\sim250\,{\rm K}$ that we obtained is in general agreement with that predicted from the difference in \ion{H}{$\beta$} equivalent width. Both averaged spectra are fitted with a surface gravity that is consistent with main sequence stars and a metallicity $[{\rm M/H}]$ that is in good agreement with photometric estimates of the cluster metallicity ($[{\rm Fe/H}]=-0.7$, e.g. \citealt{Martocchia2017}).

A rotation velocity was finally obtained by applying rotational broadening according to the formula of \citet{Gray2008}\footnote{using the implementation in \url{https://github.com/sczesla/PyAstronomy}.} to the templates in a least squares fit to the data. This yielded values of $V\sin i=87\pm16\,{\rm km\,s^{-1}}$ and $130\pm22\,{\rm km\,s^{-1}}$ for the blue and red MSTO spectrum, respectively. The uncertainies were obtained by re-noising the spectra using the uncertainties obtained as outlined above and re-evaluating $V\sin i$.

Our analysis hence provides evidence for rotational broadening of both spectra and tentative evidence for an increase of the average $V\sin i$ towards the red side of the MSTO. To visualize the effect of the measured rotational broadening on MUSE spectra, we show in the lower panel of Fig.~\ref{fig:msto_spectra} two synthetic PHOENIX spectra. They have similar stellar parameters as obtained for the combined MSTO spectra and were broadened using the $V\sin i$ values obtained for the red and blue MSTO spectrum, respectively. The synthetic spectra were transformed to the MUSE spectral resolution and sampling. As was the case for the observed spectra depicted in the upper panel, a visual comparison does not reveal a stronger rotational broadening of the red template spectrum. Hence, to reveal stellar rotation at the spectral resolution of our data, we have to rely on an in-depth analysis as performed above. To further verify our $V\sin i$ values, we performed 1\,000 realisations in which we degraded the synthetic spectra shown in the lower panel of Fig.~\ref{fig:msto_spectra} to a random S/N between 1 and 200 and attempted to recover the input values of $V\sin i$. We found that above a S/N of 50, we are able to recover the input values to an accuracy within the confidence intervals quoted above.

While this simulation gives us an idea of the accuracy that we can achieve with MUSE spectra if a perfectly matching template is available, it does not account for potential mismatches between the observed spectra and the syntetic templates. We investigated the impact of such mismatches on our analysis by modifying the analysis on the synthetic templates such that the spectra used to recover $V\sin i$ were offset from the broadened templates in effective temperature (by $200\,{\rm K}$), surface gravity (by $0.5\,{\rm dex}$), and alpha-element abundance (by $0.4\,{\rm dex}$). We found that such mismatches can indeed have a significant impact on the recovery of $V\sin i$. The offsets we measured were as large as $\pm30\,{\rm km\,s^{-1}}$. However, we did not encounter a situation were the recovered rotation for either of the spectra was consistent with zero. In addition, the systematics affected both spectra in the same sense, so that the difference in $V\sin i$ between the red and blue spectrum persisted. Hence we conclude that systematic effects have a stronger effect on the absolute values of $V\sin i$ than on their relative difference which we consider robust.


\begin{table}
	\centering
	\caption{Results of the spectral analysis of the combined MSTO spectra.} 
	\label{tab:stellar_parameters}
	\begin{tabular}{lcc} 
		\hline
		 & blue MSTO & red MSTO\\
		\hline
		$T_{\rm eff}\,[{\rm K}]$ & $7690\pm20$ & $7420\pm20$\\
		$\log g$ & $4.44\pm0.03$ & $4.41\pm0.03$\\
		$[{\rm M/H}$] & $-0.67\pm0.03$ & $-0.74\pm0.03$\\
		\hline
	\end{tabular}
\end{table}

\section{Discussion \& Conclusions}

In a study of the rotation properties of A stars in the field, \citet{royer2002a} found an unimodal distribution, peaking at equatorial velocities between $100$--$200\,{\rm km\,s^{-1}}$. While this is in general agreement with the (projected) rotation velocities we obtained in the analysis of the combined MSTO spectra, it is not obvious that A stars in clusters share the same rotation properties as their analogues in the field \citep[see, e.g.][for more massive stars]{Huang2006,Wolff2007}. An important difference between the two populations is their age distributions, as the age-evolution of the rotation of A-stars can be quite complex \citep{Zorec2012}.

Our analysis provides evidence that the stars on the red side of the main sequence rotate on average faster than their bluer counterparts, the difference in $V\sin i$ is $\sim40\,{\rm km\,s^{-1}}$. If we use the models of \citet{Georgy2014} for a 1~Gyr cluster and assume an isotropic distribution of inclination angles, we obtain a difference of $\sim 100\,{\rm km\,s^{-1}}$. Our recent work on the Galactic open cluster NGC~2818 \citep{Bastian2018} has shown that these models accurately describe the relation between colour and $V\sin i$ across the MSTO of this 0.8~Gyr old cluster. However, as 1~Gyr is the maximum age currently covered in the models and NGC~419 is still 0.5~Gyr older, there are no models that can we can directly compare to our results without extrapolation to older ages. It would be an important step forward to extend the evolutionary models to the age of NGC~419.

Finally, our analysis of the cluster dynamics yielded a dynamical mass-to-light ratio of $0.67\pm0.08$, slightly higher than the predictions of simple stellar population models and hints to the presence of rotation inside the cluster. The observation that clusters at all ages show rotation \citep[e.g.][]{Davies2011q,Henault-Brunet2012,Kamann2018} suggests that they form with significant angular momentum. A survey that traces the evolution of angular momentum across all cluster ages therefore promises to reveal crucial information about the formation of massive clusters. Our analysis of NGC~419 has shown that the combination of MUSE and adaptive optics provides us with a powerful opportunity to do so.

\section*{Acknowledgements}

We thank Roland Bacon, Joel Vernet, the MUSE team, and the staff at ESO for their efforts during the instrument commissioning and for making the data available to the community.
SK, NB, and CU gratefully acknowledge funding from a European Research Council consolidator grant (ERC-CoG-646928-Multi-Pop).
AK und PMW acknowledge funding from BMBF-Verbundforschung (grant 05A17BAA).
This work made use of PyAstronomy.




\bibliographystyle{mnras}
\bibliography{ngc419_ao_commissioning}



%
%


\bsp	
\label{lastpage}
\end{document}